\documentclass[superscriptaddress,longbibliography,twocolumn,amsmath,amssymb, footinbib]{revtex4-2}
\usepackage{tikz}
\usepackage{pgfplots}
\usepackage{graphicx}
\usepackage{dcolumn}
\usepackage{bm}
\usepackage{esint}
\usepackage{color}

\usepackage{hyperref}

\begin{document}


\title{
Large time and long distance asymptotics of the thermal correlators\\ of the impenetrable anyonic lattice gas
}

\author{Yuri Zhuravlev}
\affiliation{Bogolyubov Institute for Theoretical Physics, 03143 Kyiv, Ukraine}

\author{Eduard Naichuk}
\affiliation{Kyiv Academic University, 03142 Kyiv, Ukraine}

\author{Nikolai Iorgov}%
\affiliation{Bogolyubov Institute for Theoretical Physics, 03143 Kyiv, Ukraine}
\affiliation{Kyiv Academic University, 03142 Kyiv, Ukraine}

\author{Oleksandr Gamayun}
\email[Correspondence to: ]{oleksandr.gamayun@fuw.edu.pl}
\affiliation{Bogolyubov Institute for Theoretical Physics, 03143 Kyiv, Ukraine}
\affiliation{
Faculty of Physics, University of Warsaw, ul. Pasteura 5, 02-093 Warsaw, Poland
}%

\date{\today}

\begin{abstract}
We study thermal correlation functions of the one-dimensional impenetrable lattice anyons. These correlation functions can be presented as a difference of two Fredholm determinants. 
To describe large time and long distance behavior of these objects we use the effective form factor approach. 
The asymptotic behavior is different in the space-like and time-like regions. 
In particular, in the time-like region we observe the additional power factor on top of the exponential decay. We argue that this result is universal as it is related to the discontinuous behavior of the phase shift function of the effective fermions. At particular values of the anyonic parameter, we recover asymptotics of spin-spin correlation functions in  XXO quantum chain. 
\end{abstract}

\maketitle


\section{Introduction}

Quantum one-dimensional models always attracted a lot of attention due to rich structures 
of their correlation functions and the possibility to address non-perturbative phenomena 
\cite{korepin_bogoliubov_izergin_1993,Tsvelik_2003,Giamarchi_2003}. 
For low temperatures the culmination of these developments resulted in the formulation of effective fields theories (Luttinger model) \cite{Giamarchi_2003,RevModPhys.83.1405}. 
With the advancement of the experimental techniques in cold atom experiments \cite{Bloch_2005,Kinoshita2006,Hofferberth_2007,RevModPhys.83.863} the interest to the non-equilibrium dynamics
or dynamics of highly excited states motivated a lot of theoretical research resulting in new concepts like generalized Gibbs ensembles, the quench action 
\cite{Caux_2013,Caux_2016},
the generalized hydrodynamics \cite{Castro_Alvaredo_2016,PhysRevLett.117.207201,denardis2021correlation} (GHD) and others. 
The main approach to the correlation function in integrable models is a direct summation of the form factors in the spectral expansion. 
The computation of the correlation functions on the finite entropy states is very different from the vacuum case due to different decay rate of the form factors with the system size (exponential vs power-law).
Therefore, different approaches were developed to tackle this kind of problems including Quantum Transfer Matrix approach \cite{Dugave_2013,Dugave_2014,10.21468/SciPostPhysLectNotes.16,Ghmann_2017,Ghmann2020,Ghmann2019a1,Ghmann2020l},
non-linear differential equations \cite{Its_1990,TPrice}, 
axiomatic definition of the thermal form factors in the Integrable Quantum Field Theories \cite{LECLAIR1999624,SALEUR2000602,Mussardo_2001,Castro_Alvaredo_2002,Doyon_2005,Altshuler2006,Doyon2007,Essler2009aa,Pozsgay2010},
adaptation of the GHD methods \cite{10.21468/SciPostPhys.10.5.116,CortsCubero2019,CortesCubero2020,Cubero1}, as well as 
partial summations of the few particle-hole excitations \cite{DeNardis2015,SciPostPhys.1.2.015,DeNardis2018,Panfil2020,Panfil_2021} and extracting the most singular parts of the form factors 
\cite{10.21468/SciPostPhys.9.3.033,Etienne2}. 

Recently we have developed a method to deal with correlation functions in finite entropy states \cite{GIZ}.
This method allows one to derive behaviour of the correlation functions in free-fermionic models for the observables 
that can be expressed as Fredholm determinants of integrable kernels. In Ref. \cite{GIZ} we focused mostly on static correlation functions and have applied the method to XY  quantum chain.

In this work we continue development of the method of effective form factors for dynamical correlation functions.
As a model of interest we choose one-dimensional impenetrable anyons on a lattice \cite{Patu2015}.
This model describes quantum particles with unusual statistics \cite{Patu2007,Patu2008,Patu2008a,Patu2009,Patu2015,Zinner}, which can be realized 
experimentally in ultracold quantum gases confined in optical traps \cite{Keilmann2011, PhysRevA.97.053605, PhysRevA.94.023615, PhysRevLett.117.205303, PhysRevLett.115.053002, PhysRevLett.101.260501, PhysRevLett.91.090402, Micheli2006, Jiang2008}. 
Furthermore,  this type of models appears after the spin-charge separation in interacting systems  of spinful  fermions 
and spin chains (at certain values of the anyonic parameter) \cite{Berkovich1987,Berkovich_1991,Ghmann1998,IZERGIN1998594,Ghmann1998a,PhysRevLett.92.176401,PhysRevLett.93.226401,Cheianov_2004,PhysRevA.100.063635}.
Similar determinants also can be obtained as the correlation functions of  Wigner strings \cite{tierrygiamarchi}. 
Also they appear in the describing the mobile impurity propagating in the gas of free fermions \cite{Gamayun_2016,GAMAYUN201583,PhysRevLett.120.220605,10.21468/SciPostPhys.8.4.053}. 
In the latter case the anyonic parameter can be identified with the total momentum of the system (at the infinite coupling). 

The main idea of the effective form factor approach is to replace computation of the correlation functions averaged over some ensemble 
to zero temperature correlators with the appropriately modified  phase shift. 
The correlation functions for one-dimensional impenetrable anyons can be presented as a linear combination of the Fredholm determinants \cite{Patu2015}.
Therefore we may identify the phase shift comparing these determinants to the one that emerges from the summation of the effective form factors.
For the space-like region we can simplify the corresponding kernels for large time and space separation and find the effective phase shift for all values of the quasi momenta. 
The time-like region is characterized by the critical points that separate different types of the asymptotic behavior. So we can robustly find the effective phase shift 
only away from these points. Even though the vicinity of critical points where we do not know the solutions vanish in the large time limit, we cannot simply combine solutions in the different asymptotic regions 
into a single phase shift as the latter will be discontinuous. To tackle this problem we have assumed the existence of the gluing regularization functions. 
While we have not been able to find them explicitly, we have demonstrated that they only affect the overall constant in the asymptotic expression of the Fredholm determinants. 

The structure of the paper is as follows. In Sec.~\ref{sect:model} we define the anyonic model and review important results about this model, such as spectrum and presentation of correlation functions in terms of Fredholm determinants. In Sec.~\ref{sect:EFFA} we recall the effective form factor approach and give two expressions for the $\tau$ function in the thermodynamic limit.    
In Sec.~\ref{sect:asymp} the  effective form factor approach  is applied to the derivation of the large time and long distance asymptotics of the dynamical correlation functions. We discuss separately  space-like and time-like regimes. In Sec.~\ref{sect:summary} we summarize the main results of the paper, compare with the known results in literature and discuss different possibilities for further research.
Appendix contains technical details of the asymptotic analysis of the form factors with the regularized effective phase shift.

\section{Model}\label{sect:model}

The one-dimensional impenetrable lattice anyons on $L$ sites can be described by the following Hamiltonian \cite{Patu2015} 
\begin{gather}\label{H}
    H=-\sum_{j=1}^{L}\frac{1}{2}(a^{\dag}_j a_{j+1}+a^{\dag}_{j+1} a_j)+h\sum_{j=1}^{L}a^{\dag}_j a_j, \\ 
    a_{L+1}=a_1, \qquad a^{\dag}_{L+1}=a^{\dag}_1.
\end{gather}
The operator algebra is specified by the anyonic parameter $0\le \kappa \le 1$ and reads as  
\begin{subequations}
\begin{gather}
a_j a^{\dag}_k=\delta_{jk}-e^{-i\pi\kappa\epsilon(j-k)} a^{\dag}_k a_j, \\
a_j a_k=-e^{i\pi\kappa\epsilon(j-k)}a_k a_j, \\
a^{\dag}_j a^{\dag}_k=-e^{i\pi\kappa\epsilon(j-k)}a^{\dag}_k a^{\dag}_j,
\end{gather}
\end{subequations}
here  $\epsilon(j)=\mathrm{sign}(j)$ and we prescribe that $\epsilon(0) = 0$. 

The case $\kappa=0$ corresponds to fermions, and $\kappa=1$ describes operators in the Hilbert space of the impenetrable bosons. 
Note, also that in the latter case the Hamiltonian \eqref{H} 
can be identified with the Hamiltonian of quantum XX spin chain after the mapping
$a_j=\sigma_j^+$, $a_j^\dag=\sigma_j^-$.

Spectrum of the Hamiltonian $H$ can be found by means of Bethe ansatz. 
The $N$-particle states are labeled by $N$ momenta $\{p_1,p_2,\ldots,p_N\}$ from the set of $L$ inequivalent 
solutions of the equation 
\begin{equation}
e^{ipL}=e^{-i\pi\kappa(N-1)}.
\end{equation}
The energies of such states are
\begin{gather}
    E(\{p_1,p_2,\ldots,p_N\})=\sum_{j=1}^{N}\varepsilon(p_j), \\
    \varepsilon(p)=h-\cos p.
\end{gather}

An interesting and non-trivial problem in the considered model is to analyze 
two-point correlation functions 
\begin{equation}\label{gmdef}
G_{-}(x,t)=\frac{\mathrm{Tr}[e^{-\beta H}a^{\dag}_{x+1}(t)a_{1}(0)]}{\mathrm{Tr}[e^{-\beta H}]},
\end{equation}
\begin{equation}
G_{+}(x,t)=\frac{\mathrm{Tr}[e^{-\beta H}a_{x+1}(t)a^{\dag}_{1}(0)]}{\mathrm{Tr}[e^{-\beta H}]}.
\end{equation}
It is easy to check the symmetry relations
\begin{equation}
    G_\pm(-x,-t)=G_\pm(x,t)^*, 
\end{equation}
and also for $t=0$
\begin{equation}
    G_{-}(x,0)+e^{-i\pi\kappa\,\mathrm{sign}(x)}G_{+}(-x,0)=\delta_{x,0}
\end{equation}
which allow us to consider only $t\ge 0$. 
In what follows we will restrict ourselves to the analysis of correlator $G_{-}(x,t)$. An analogous analysis can be done for  $G_{+}(x,t)$.
It was shown that these correlators  in the thermodynamic limit $L\to\infty$ can be written in terms of
Fredholm determinants \cite{Patu2015}. 
We will use the following equivalent presentation for $G_{-}(x,t)$:
\begin{equation}\label{gmdet}
G_{-}(x,t)=\det(1+\hat W+\delta \hat W)-\det(1+\hat W),
\end{equation}
where  $\hat W$ and  $\delta \hat W$ are integral operators on $[-\pi,\pi]$ with the kernels 
\begin{gather}\label{gm1}
W(p,q)=\frac{1}{2\pi}e_{-}(p)e_{-}(q)e^{\frac{i(p-q)}{2}}\frac{e(p)-e(q)}{\sin\frac{p-q}{2}}, \\ 
\delta W(p,q)=\frac{1}{2\pi}e_{-}(p)e_{-}(q),\\
n_F(p)=\frac{1}{e^{\beta\varepsilon(p)}+1},\\
e_{-}(p)=\sqrt{n_F(p)}e^{-ixp/2+it\varepsilon(p)/2},
\end{gather}
\begin{multline}
e(p)=\sin^2\frac{\pi\kappa}{2} \fint_{-\pi}^{\pi}\frac{dq}{2\pi}e^{ixq-it\varepsilon(q)}\cot\frac{q-p}{2}\\
+\frac{1}{2} \sin (\pi\kappa) e^{ixp-it\varepsilon(p)}.
\end{multline}
Eq.~\eqref{gmdet} allows us to compute the correlation function $G_{-}(x,t)$ numerically.
However, large time and long distance asymptotics of the correlation functions are hard to extract  by numerical means due to the oscillatory behavior of integral kernels. In the present paper we instead analyze these asymptotics analytically by means of the effective form factor approach \cite{GIZ}, which is briefly reviewed below.  

\section{Effective form factor approach}\label{sect:EFFA}
\subsection{Effective form factors and  tau function}

In this section we recall the effective form factor approach initiated in \cite{GIZ}. 
To specify the effective form factor we require two smooth periodic functions $\nu(k)$, $g(k)$. 
The first one is called the effective phase shift and defines the shifted set of momenta as solutions of
\begin{equation}\label{eqk}
    e^{ikL} = e^{-2\pi i \nu(k)}.
\end{equation}
Here $L$ is regarded as a system size.
Since $\nu(k)$ is periodic, i.e. has a zero winding number in term of \cite{GIZ}, the largest ordered set of 
the {\em shifted} momenta has $L$ terms $\mathbf{k}=\{k_1,\ldots, k_L\}$. Each $k_i$ is a solution of \eqref{eqk}. 
The {\em unshifted}  momenta are solutions of 
\begin{equation}
    e^{iqL}=1.
\end{equation}
All momenta are considered up to the equivalence  ${k\sim k+2\pi}$ and it is convenient to choose them to have real parts in  the Brillouin zone $[-\pi,\pi]$.

The effective form factors are defined for the subsets of momenta $\mathbf{q}$ of the size $L-1$. Such subsets can be parameterized by the position of the ``hole''
\begin{equation}\label{qaFF}
    \mathbf{q}^{(a)}=\{q_1,\ldots,q_{a-1},q_{a+1},\ldots,q_L\},\quad a=1,\ldots,L.
\end{equation}
The effective form factor then reads 
\begin{multline}
 |\langle \textbf{k}|\mathbf{q}^{(a)}\rangle|^2=
 L^{1-2L}\prod_{j=1}^L \frac{e^{g(k_j)-g(q_j)} \sin^2\pi\nu(k_j)}{1+\frac{2\pi}{L}\nu'(k_j)}\\
 \times
 e^{g(q_a)} {\det}^2 D^{a},
\end{multline}
where ${\det} D^{a}$ is defined for $\mathbf{q}^{(a)}$ and is nothing but a trigonometric variation of the Cauchy determinant, in which the row corresponding to $q_a$ is omitted and replaced with the line of $1$
\begin{equation}
{\det} D^{a} = \begin{vmatrix}
\cot\frac{k_1-q_1}{2} &\dots & \cot\frac{k_{L}-q_1}{2}\\
\vdots & \ddots & \vdots \\
\cot\frac{k_1-q_L}{2} &\dots & \cot\frac{k_{L}-q_L}{2}\\
1& \dots  & 1\\
\end{vmatrix},
\end{equation}
 as we deal only with the square of the determinant we can put this line to be the last one. 
 
The tau (correlation) function is defined as series over these form factors 
\begin{equation}\label{tau0}
\tau(x,t)= \sum_{{\bf q}^a} |\langle {\bf k}|{\bf q}^a\rangle|^2  e^{-ix (P({\bf k})-P({\bf q}^a))+it(E({\bf k})-E({\bf q}^a))}.
\end{equation}
Here we use notations for the momentum and energy of many-particle state $|{\bf q}\rangle$
\begin{equation}
    P({\bf q})=\sum_{q\in {\bf q}} q,\quad
    E({\bf q})=\sum_{q\in {\bf q}} \varepsilon(q).
\end{equation}
In Ref.~\cite{GIZ} we have demonstrated that in the thermodynamic limit $L\to \infty$ the tau function can be presented  as a difference of two Fredholm determinants
\begin{equation}\label{taufred}
\tau(x,t)=\det(1+\hat V+\delta \hat V)-\det(1+\hat V),
\end{equation} 
where $\hat V$ and $\delta \hat  V$ are integral operators on $[-\pi,\pi]$ with kernels
\begin{gather}
V(p,q)=\frac{1}{2\pi}
c_-(p)c_-(q) e^{\frac{i(p-q)}{2}}\frac{c(p)-c(q)}{\sin\frac{p-q}{2}},\\
\delta V(p,q)=\frac{1}{2\pi}c_-(p)c_-(q),
\end{gather}
\begin{equation}
    c_{-}(p)=\sin\pi\nu(p) e^{-ixp/2+it\varepsilon(p)/2+g(p)/2},
\end{equation}
\begin{multline}
c(p)=\fint_{-\pi}^{\pi}\frac{dq}{2\pi}e^{ixq-it\varepsilon(q)-g(q)}\cot\frac{q-p}{2}\\
+\cot\pi\nu(p)e^{ixp-it\varepsilon(p)-g(p)}.
\end{multline}
This form  allows  us to relate the correlation function of anyons with the tau function for a special choice of $\nu(k)$ and $g(k)$. This relation will be described in the next section.

\subsection{Finite size scaling}

In this subsection we give an alternative formula for the tau function 
based on the first taking the thermodynamic limit of the form factors and then performing the summation. 
The obtained expressions will have a simple form convenient for asymptotic analysis \cite{GIZ}.

We start from representing ${\det} D^{a}$ in a factorized form
\begin{equation}\label{detDW1W2}
    \prod_{i=1}^L \frac{\sin^2 \pi \nu(k_i)}{L^2} {\det}^2 D^{(a)}=Z^2 \mathcal{Z}_a,
\end{equation}
\begin{gather}\label{ZZa}
    Z=   \prod\limits_{i=1}^L\prod\limits_{j=1}^{i-1} \frac{\sin\frac{k_i-k_j}{2}}{\sin\frac{q_i-q_j}{2}},\\ 
    \mathcal{Z}_a = \sin^2 \frac{\pi \nu(k_a)}{L}\prod\limits_{j\neq a}^{L} \frac{\sin^2\frac{k_j-q_a}{2}}{\sin^2\frac{q_j-q_a}{2}}.
\end{gather}
Extracting the hole dependent factors the tau function \eqref{tau0} can be rewritten as 
\begin{equation}\label{tauTL}
    \tau(x,t)=L\cdot K(x,t) \cdot \sum\limits_{a=1}^L e^{g(q_a)}\mathcal{Z}_a e^{-ixq_a+it\varepsilon(q_a)},
\end{equation}
where $K(x,t)$ is $a$-independent part given by
\begin{multline}
    K(x,t)=Z^2  e^{-ix (P({\bf k})-P({\bf q}))+it(E({\bf k})-E({\bf q}))}\\
    \times \prod_{j=1}^L\frac{e^{g(k_j)-g(q_j)}}{1+\frac{2\pi}{L}\nu'(k_j)}.
\end{multline}
The expressions $Z^2$ and $K(x,t)$ have finite thermodynamic limit \cite{GIZ}
\begin{equation}\label{Z2TL}
    \log Z = -\int\limits_{-\pi}^\pi dq \int\limits_{-\pi}^\pi dk \left[\frac{\nu(q)-\nu(k)}{4\sin \frac{q-k}{2}}\right]^2,
\end{equation}
\begin{multline}\label{Kxt}
    \log K(x,t)= 2\log Z -\int\limits_{-\pi}^{\pi}\nu(q)g'(q)dq  \\
    +i\int\limits_{-\pi}^{\pi}(x-\varepsilon'(q)t)\nu(q)dq.
\end{multline}
The hole dependent factors are suppressed in the thermodynamic limit
\begin{equation}\label{ZaTL}
    \mathcal{Z}_a\approx L^{-2}\sin^2\pi\nu(q_a)\exp\left(-\fint\limits_{-\pi}^{\pi}dq\,\nu(q)\cot\frac{q-q_a}{2}\right),
\end{equation}
but the whole tau function \eqref{tauTL} has finite thermodynamic limit and can be presented as an integral
\begin{multline}\label{tauint}
    \tau(x,t)= K(x,t) 
    \int\limits_{-\pi}^{\pi}\frac{dk}{2\pi}e^{g(k)}\sin^2\pi\nu(k)e^{-ixk+it\varepsilon(k)}\\
    \times \exp\left(-\fint\limits_{-\pi}^{\pi}dq\,\nu(q)\cot\frac{q-k}{2}\right).
\end{multline}

Thus we have two alternative presentation of tau function in the thermodynamic limit: 
Eq.~\eqref{taufred} as a difference of Fredholm determinants and Eq.~\eqref{tauint} in terms of integrals.
The first form is convenient for the identification with other models and the second form is convenient for large $x$ and $t$ analysis.

\section{Asymptotic behaviour of anyonic correlation function}\label{sect:asymp}
\subsection{Anyons and effective fermions}
To apply the method of effective form factors for the large $x$ and $t$ asymptotics of the correlation function $G_-(x,t)$ given by \eqref{gmdet}, 
we have to find suitable functions $\nu(k)$ and $g(k)$. This can be done after the identification of the kernels in \eqref{gmdet} and in \eqref{taufred}. 
In this section we focus on the case of $h>0$, the case $h<0$ can be considered similarly.
Also we restrict the value of parameter of anyonic statistics to $0\le \kappa<1$. The peculiarities with the limiting case $\kappa=1$ corresponding to quantum XX spin chain are briefly discussed in  Sec.~\ref{sect:summary}.

Equating $G_-(x,t)=\tau(x,t)$, we see that their integral  kernels coincide if we choose 
$\nu(p)$ and $g(p)$ to satisfy the equations
\begin{equation}
    c_-(p)=e_-(p), \quad c(p)=e(p).
\end{equation}
The first equation gives a relation between $g(p)$ and $\nu(p)$
\begin{equation}\label{gnu}
    e^{-g(p)}=\frac{\sin^2\pi\nu(p)}{n_F(p)}.
\end{equation}
The second equation allows us to obtain an integral equation for $\nu(p)$
\begin{multline}\label{eqnu}
\int_{-\pi}^{\pi}\frac{dq}{2\pi}\left(\frac{\lambda_+(q)}{\tan\frac{q-p-i0}{2}}+\frac{\lambda_-(q)}{\tan\frac{q-p+i0}{2}}\right)
e^{ixq-it\varepsilon(q)}=0,
\end{multline}
where we have denoted 
\begin{gather}
   \lambda_\pm(q)=\frac{e^{\pm 2i\pi\nu_\pm(q)}-e^{\pm 2i\pi\nu(q)}}{n_F(q)},
\end{gather}
\begin{gather}\label{nupm}
e^{\pm 2i\pi\nu_\pm(q)}=1+n_F(q)(e^{\pm i\pi\kappa}-1).
\end{gather}
We can solve Eq.~\eqref{eqnu} asymptotically for large $x$ and $t$.
The solution has different forms for two different values of $v\equiv x/t$. We call $|v|>1$ the space-like region and $|v|<1$ the time-like region. 
This names should not be confused with the similar in the relativistic theory --- there the spectrum is linear for all momenta. 
In our case the names come from the condition where the function 
\begin{equation}\label{Phi}
    \Phi(q) \equiv  v q+\cos q 
\end{equation}
has (time-like) or does not have (space-like) a critical point for $q \in [-\pi,\pi]$.  This function is nothing but the phase $x q - \varepsilon(q) t$ up to rescaling by time and shift 
by the constant $h$.

\subsection{Asymptotic behaviour of correlation function in space-like region} 
\label{subsect:space}

\begin{figure}
    \centering
    \includegraphics[width=0.9\linewidth]{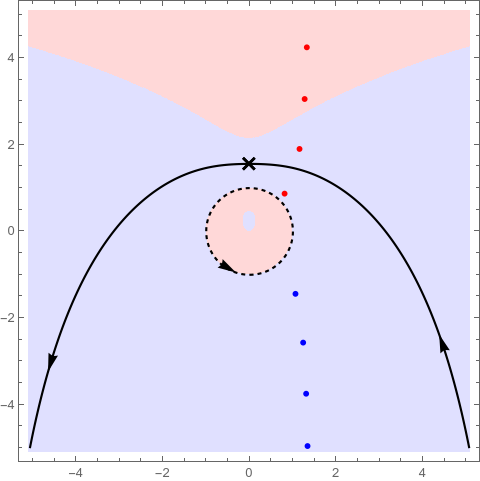}
    \caption{The integration contours (color online). The dashed circle corresponds to the initial contour of integration $C_>$.
    The black solid line represents the steepest descent contour $C_1$. The cross marks the position of the saddle point. Red and blue dots correspond to the poles $z_n$ defined by Eq.~\eqref{zn} for non-negative and negative indices, respectively.
    The shaded areas show the regions of positive (pink) and negative (light blue) values of $\mathrm{Re}\, \theta(z)$,
    see Eq.~\eqref{thetaz}.
    }
    \label{fig1}
\end{figure}

To treat Eq.~\eqref{eqnu} we first have a look on each of the integrals separately. It is useful to present them as
\begin{multline}\label{nulambda}
    \int_{-\pi}^{\pi}\frac{dq}{2\pi} \frac{\lambda_\pm(q)e^{it\Phi(q)}}{\tan\frac{q-p\mp i0}{2}}
    =    \lambda_\pm(p) \int_{-\pi}^{\pi}\frac{dq}{2\pi} \frac{e^{it\Phi(q)}}{\tan\frac{q-p\mp i0}{2}}\\+
    \int_{-\pi}^{\pi}\frac{dq}{2\pi} \frac{(\lambda_\pm(q)-\lambda_\pm(p))e^{it\Phi(q)}}{\tan\frac{q-p}{2}}.
\end{multline}
If we assume that $\nu(q)$ does not become singular even in the asymptotic region, then in space-like region the second term in RHS of Eq.~\eqref{nulambda} becomes exponentially small  
for large $x$ and $t$. The remaining integral in \eqref{nulambda} can be rewritten as
\begin{equation}\label{fpm1}
    \int_{-\pi}^{\pi}\frac{dq}{2\pi i} \frac{e^{it\Phi(q)}}{\tan\frac{q-p\mp i0}{2}}=e^{it\Phi(p)}(F(p)\pm 1),
\end{equation}
where
\begin{equation}
    F(p)=e^{-it\Phi(p)} \fint_{-\pi}^\pi \frac{dq}{2\pi i} \frac{e^{it\Phi(q)}}{\tan\frac{q-p}{2}} .
\end{equation}
For large $t>0$ the function $F(p)$ can be approximated up to exponentially small terms as
\begin{equation}
    F(p)\approx \mathrm{sign}\, (\Phi'(p)).
    \label{crf}
\end{equation}
As the space-like region is characterized by the absence of critical points of $\Phi(p)$ for $p\in [-\pi,\pi)$, we can put  ${\mathrm{sign}\, (\Phi'(p))=\mathrm{sign}\, v}$. 
This way, one of the two integrals in Eq.~\eqref{eqnu}  is exponentially small due to \eqref{nulambda} and \eqref{fpm1}, while the other allows us to find the effective phase shift for large $t>0$
\begin{equation}\label{nuspace}
    \nu(p)\approx \nu_{\mathrm{sign}\, v}(p),
\end{equation}
where $\nu_\pm(p)$ are defined by Eqs.~\eqref{nupm}.
We use this asymptotic solution and the relation \eqref{gnu} in \eqref{tauint} to obtain 
\begin{equation}\label{tauintspace}
    \tau(x,t)= K(x,t) T(x,t) e^{ith},
\end{equation}
where $K(x,t)$ is given by  Eq.~\eqref{Kxt} and $T(x,t)$ corresponds to the integral in  \eqref{tauint}, which after the change of variables ${z=e^{ik}}$ takes the following form
\begin{equation}
    T(x,t)=\frac{1}{2\pi i}\oint_{C_{>}}\frac{dz}{z}\frac{e^{t\theta(z)}S(z)}{J(z)+e^{i\pi\kappa}}.
\end{equation}
Here $C_{>}$ is a counterclockwise circle with a radius slightly larger than $1$ and 
\begin{equation}\label{thetaz}
    \theta(z)=-v\log z-\frac{i}{2}(z+z^{-1}),
\end{equation}
\begin{equation}
     S(z)=\exp\left( i\int\limits_{-\pi}^{\pi} dq\, \nu(q)  \frac{z+e^{iq}}{z-e^{iq}}\right),
\end{equation}
\begin{equation}
    J(z)=\exp\left(\beta\left(h-\frac{z+z^{-1}}{2}\right)\right).
\end{equation}
In what follows we consider $v>1$, the other case,  ${v<-1}$, can be considered in the same manner. 
In order to find large $x$ and $t$ asymptotics of  $T(x,t)$ we deform the contour $C_>$ to the steepest descend curve $C_1$
defined by
\begin{equation}
    \mathrm{Im}\,\theta(z)=\mathrm{Im}\,\theta(z_\mathrm{sp})=-\pi v/2
\end{equation}
going through the saddle point $z_\mathrm{sp}$ 
\begin{equation}
    z_\mathrm{sp}=iv+i\sqrt{v^2-1}.
\end{equation}
Deforming the contour we might cross the poles of integrand, which can only appear from the denominator, since $S(z)$  is a holomorphic function for $|z|>1$. This way, we get
\begin{multline}\label{TxtC}
    T(x,t)=\frac{1}{2\pi i}\oint_{C_{1}}\frac{dz}{z}\frac{e^{t\theta(z)}S(z)}{J(z)+e^{i\pi\kappa}} \\ -\sum_{n=-\infty}^{n_0}\mathrm{res}_{z=z_n}\frac{e^{t\theta(z)}S(z)}{z(J(z)+e^{i\pi\kappa})},
\end{multline}
where the points $z_n$ are defined as
\begin{equation}\label{zn}
z_n=h_n+\sqrt{h_n^2-1}, \ \  h_n=h+\frac{i\pi}{\beta}(2n+1-\kappa),
\end{equation}
and $n_0$ is the maximal number of a pole, which was crossed in the deformation process. This number depends on the velocity $v$ and can be found from the inequality
\begin{equation}
    \mathrm{arg}\,z_{n_0} < \frac{\pi}{2}-\frac{h}{v}<\mathrm{arg}\, z_{n_0+1}.
\end{equation}
Schematically the contours $C_>$, $C_1$ and the positions of poles $z_n$ are  shown in Fig.~\ref{fig1}.

The formula \eqref{TxtC} allows one immediately to read off the asymptotic behaviour. 
The residues produce exponentially decaying terms; the leading contribution is given by the smallest real part 
$\mathrm{Re}\, \theta(z_n)$. For wide range of the parameters of the model we observed that this achieved for the pole at $z_0$. 
Another type of contribution to the asymptotics comes from the saddle point evaluation of the integral in~\eqref{TxtC}.
To find the overall leading contribution we need to compare $\mathrm{Re}\, \theta(z_0)$ and
$\mathrm{Re}\, \theta(z_\mathrm{sp})$. 
This leads to the equation for the critical velocity $v_c$ separating two regimes
\begin{equation}
    -v_c \log(v_c + \sqrt{v_c^2-1}) + \sqrt{v_c^2-1} = - v_c \log|z_0| + \frac{\pi}{\beta} (1-\kappa).
\end{equation}

\begin{figure}
    \centering
    \includegraphics[width=\linewidth]{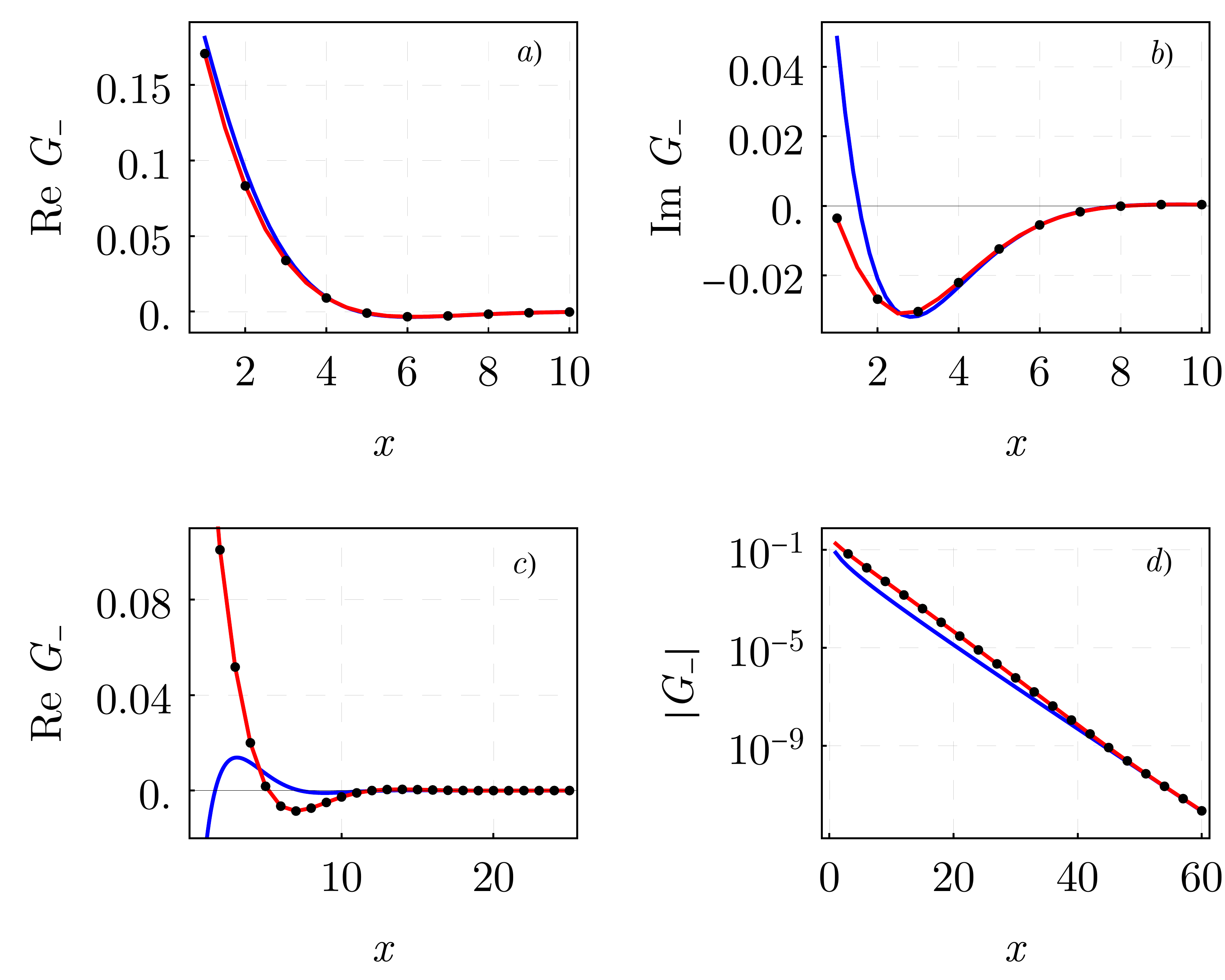}
    \caption{Asymptotic behavior  of $G_-(x,t)$ for ${\kappa=0.6}$, $h=0.7$, $\beta=2.3$. These parameters correspond to critical velocity $v_c\approx 1.676$.
    Black dots present $G_-(x,t)$ computed numerically from \eqref{gmdet}.     
    Red lines present effective $\tau$ function \eqref{tauintspace} computed with \eqref{nuspace}.  Blue lines present asymptotics of
    integrals in \eqref{tauintspace} given by Eqs.~\eqref{txtsp}, \eqref{txtpole}.  Panels $a)$ and $b)$ correspond to overcritical region $v=2.5$. Panels $c)$ and $d)$ show real part and absolute value of $G_-(x,t)$ in the subcritical region $v=1.3$, respectively.}
    \label{figspacesaddle}
\end{figure}

For $v<v_c$ the saddle point is dominating and $T(x,t)$ is given by
\begin{equation}\label{txtsp}
    T(x,t) \approx \frac{S(z_{\rm sp}) }{J(z_{\rm sp})+e^{i\pi\kappa}} \frac{e^{ -x\log z_{\rm sp} - \frac{it}{2}(z_{\rm sp}+z_{\rm sp}^{-1})}}{\sqrt{2\pi t\sqrt{v^2-1}}},
\end{equation}
for $v\ge v_c$ the pole gives the leading contribution 
\begin{equation}\label{txtpole}
    T(x,t) \approx- \frac{2}{\beta} \frac{e^{-i \pi \kappa}}{z_0-z_0^{-1}} S(z_0) e^{ -x\log z_0 - \frac{it}{2}(z_0+z_0^{-1})}.
\end{equation}
The asymptotics for the tau function can be found from Eq.~\eqref{tauintspace}, for which we also provide the simplified expression for $K(x,t)$
\begin{equation}
    \log K(x,t) = 2\log Z +ix \int\limits_{-\pi}^{\pi}\nu(q)dq,
\end{equation}
where $\nu(q)$ is given by Eqs.~\eqref{nuspace} and \eqref{nupm}.
We compare these asymptotic expressions for the correlation functions with numerical evaluation of Fredholm determinants 
\eqref{gmdet} in Fig.~\ref{figspacesaddle}. We see that asymptotics given by the integral (the red solid line), i.e. by the tau function is hardly distinguishable from the 
true correlation function even for small $x$.

\subsection{Asymptotic behaviour of correlation function in time-like region}

Now let us try to apply the same reasoning for the time-like region, $|v|<1$. In this case there are two critical points $q_1$ and $q_2$
\begin{equation}
    \Phi'(q_i)=0,\qquad q_i \in [-\pi,\pi),
\end{equation}
therefore the approximation \eqref{crf} naively gives rise to the solution 
\begin{equation}\label{nutime}
    \nu(p)\approx \nu_{\mathrm{sign}\, \Phi'(p)}(p),
\end{equation}
where $\nu_\pm(p)$ are defined by Eqs.~\eqref{nupm}. This is
valid for all $p$ lying far enough from the critical points. Indeed, the approximation \eqref{crf} holds everywhere outside small vicinities of width $\sim t^{-1/2}$ around critical points $q_1$ and $q_2$. 

It is very tempting to ignore these domains and approximate $\nu(p)$ as a truly discontinuous function, since we are interested in the large $t$ behavior. This procedure however is not consistent with the approximations made in Eq.~\eqref{nulambda} where we have discarded critical point contributions (the last integral). But even bigger problems appear when one tries to use discontinuous $\nu(p)$ for the asymptotic expression. For instance the double integral \eqref{Z2TL} is divergent for such a choice. 

\begin{figure}[b]
    \centering
    \includegraphics[width=0.9\linewidth]{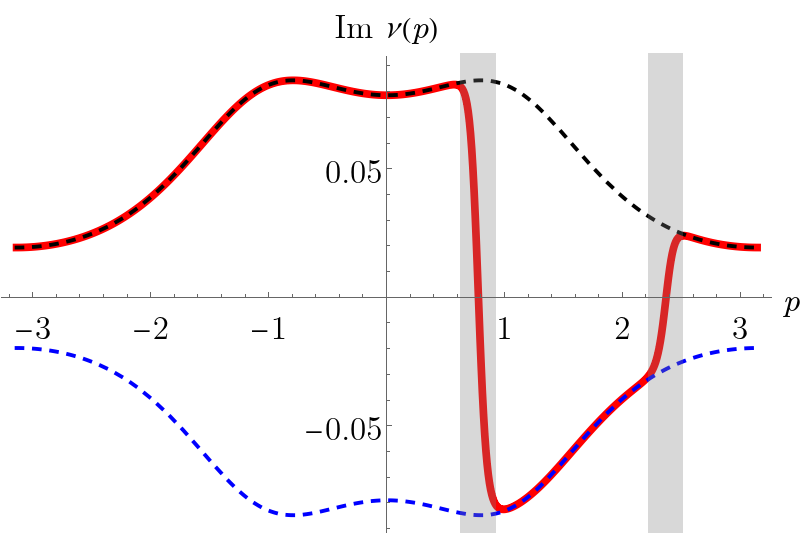}
    \caption{The schematic dependence of the effective phase shift $\nu(p)$. 
    The black and blue dotted lines represent $\nu_+(q)$ and $\nu_-(q)$, respectively. 
    The red lines shows the regularized expression for $\nu(p)$. The shaded rectangles show the regions where the transition between $\nu_+$ and $\nu_-$ happens and the regularization is required to  approximate $\nu(p)$. These regions are located near critical points $q_1, q_2$ and their widths are $O(t^{-1/2})$. We show only the imaginary part as the real part is continuous and does not require regularization.
    }
    \label{fignu}
\end{figure}

Therefore, we expect that the solution of the Eq. \eqref{eqnu} will have the following ``regularized'' form
\begin{equation}\label{nuR}
    \nu(p)=A(p)+B(p)s(p), 
\end{equation}
where
\begin{equation}
    A(p)=\frac{\nu_+(p)+\nu_-(p)}{2}, \qquad B(k)=\frac{\nu_+(p)-\nu_-(p)}{2},
\end{equation}
and the function $s(p)$ is a regularization of $\operatorname{sign}$ function
\begin{equation}
    s(p)=f(\sqrt{t}\Phi'(p)), 
\end{equation}
with $f$ being a smooth function satisfying
\begin{equation}\label{eqf1}
      \quad f(\pm\infty)=\pm 1.
\end{equation}
So away from the critical points on the distance bigger than $O(1/\sqrt{t})$ we recover the solution \eqref{nutime}. 
We demonstrate this schematically in Fig.~\ref{fignu}. Notice that the regularization is needed only for the imaginary parts and the real parts of $\nu_+(p)$ and $\nu_-(p)$ coincide. 
Now for the smooth $\nu(p)$ we can use all the results from the previous sections. In particular, we can integrate Eq.~\eqref{Z2TL} 
by parts to obtain 
\begin{equation}
   \log Z=\frac{1}{2}\int\limits_{-\pi}^{\pi}dq\int\limits_{-\pi}^{\pi}dk\, \nu'(q)\nu'(k)\log\left|\sin\frac{q-k}{2}\right|.
\end{equation}
We can perform asymptotic analysis of this expression for large $t$ and obtain 
\begin{equation}
    Z\approx t^{-\frac12 (\delta_1^2+\delta_2^2)}\,Z_\mathrm{reg} ,
\end{equation}
where $Z_\mathrm{reg}$ is $t$-independent factor depending on $s(p)$ and 
\begin{equation}
    \delta_1=\nu_-(q_1)-\nu_+(q_1),\quad
    \delta_2=\nu_+(q_2)-\nu_-(q_2).
\end{equation}
Therefore the only regularization dependence remains in the overall constant prefactor. 
It is remarkable that the exponent of power law $t$-dependence of $Z$ is universal (does not depend on the regularization $s(p)$ for any $f$ satisfying \eqref{eqf1}). 
These computations and the exact form for $Z_\mathrm{reg}$ are given in the Appendix. 

Let us also discuss the asymptotic behavior of the remaining part of the tau function. In there we substitute already discontinuous $\nu(q)$. Namely, we analyze the integral
\begin{equation}
\label{Txt}
    T(x,t)=  
    \int\limits_{-\pi}^{\pi}\frac{dk}{2\pi}n_F(k)e^{-it\Phi(k)}e^{-Y(k)},
\end{equation}
where
\begin{equation}
    Y(k)=\fint\limits_{-\pi}^{\pi}dq\,\nu(q)\cot\frac{q-k}{2}.
\end{equation}
The function $Y(k)$ is logarithmically divergent at $q_1$ and $q_2$ because of discontinuity of $\nu(q)$.
It leads to power like singularities in the integrand of \eqref{Txt} which are integrable if $\mathrm{Re}\, \delta_j>-\frac12$.
In our case $\nu_+(k)$ and $\nu_-(k)$ are conjugate to each other rendering the real part of effective phase shift to be  continuous,  
$\mathrm{Re}\, \delta_j=0$.

We separate a regular part $\tilde{Y}(k)$ of $Y(k)$ as
\begin{equation}
    Y(k)
    =\tilde Y(k)+ (\nu_-(k)-\nu_+(k)) \log\left(  \frac{\sin\frac{q_1-k}{2}}{\sin\frac{q_2-k}{2}}\right)^2,   
\end{equation}
\begin{multline}\label{tildeS}
    \tilde{Y}(k)=\int\limits_{-\pi}^{q_1}dq\,(\nu_+(q)-\nu_+(k))\cot\frac{q-k}{2}\\
    +\int\limits_{q_1}^{q_2}dq\,(\nu_-(q)-\nu_-(k))\cot\frac{q-k}{2}\\
    +\int\limits_{q_2}^{\pi}dq\,(\nu_+(q)-\nu_+(k))\cot\frac{q-k}{2}.
\end{multline}
Now all is prepared to find the asymptotic behaviour of $T(x,t)$ for large $x$ and $t$ coming from the contributions of two critical points $q_1$ and $q_2$
\begin{equation}\label{Txtasymp}
    T(x,t)\approx T_1+T_2, 
\end{equation}
where 
\begin{multline}
T_j=\frac{n_F(q_j)}{2\pi} e^{-it\Phi(q_j)-\tilde Y(q_j)} \left( 2\sin\frac{q_2-q_1}{2} \right)^{-2\delta_j}  \\
\times \left(\frac{it\Phi''(q_j)}{2}\right)^{-\frac12 - \delta_j} \Gamma\left(\frac12 +\delta_j\right).
\end{multline}

\begin{figure}
    \centering
    \includegraphics[width=\linewidth]{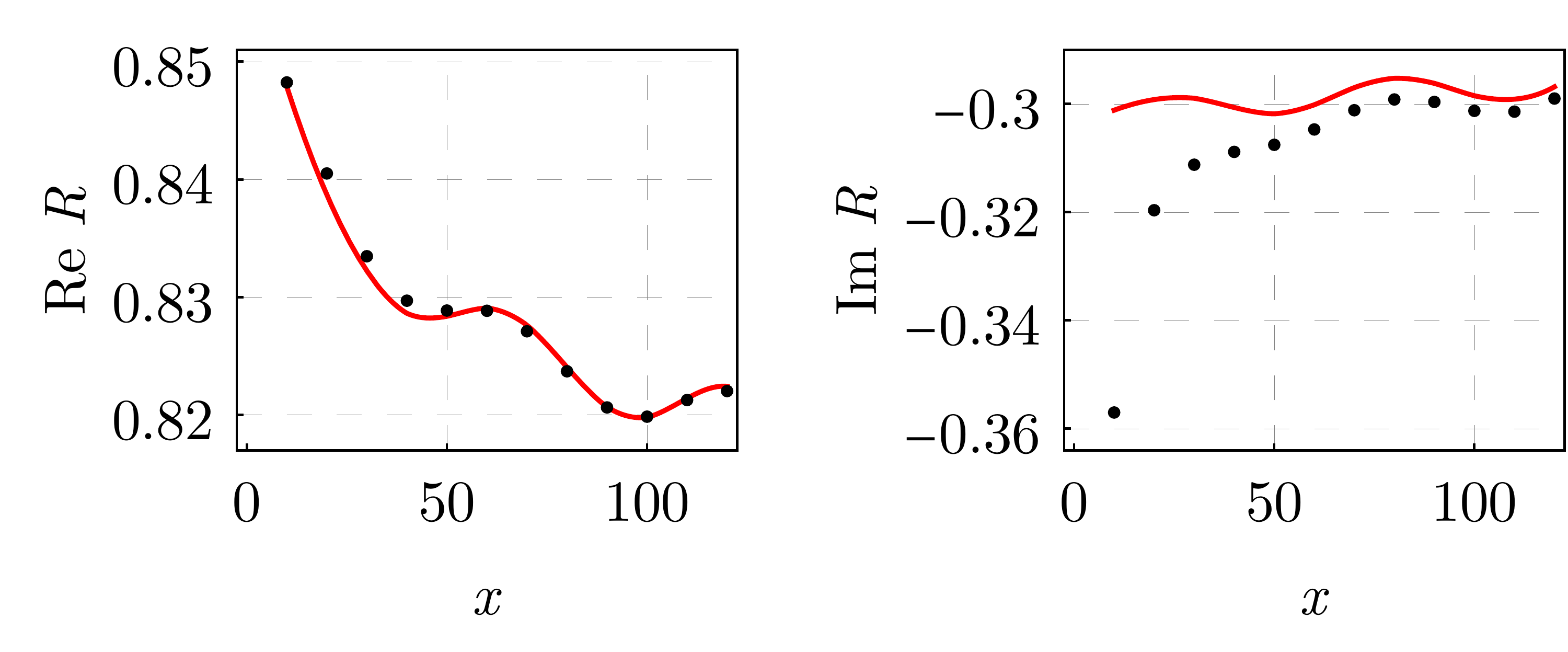}
    \caption{Real and imaginary part of $R(x,t)$ with $v=x/t=0.5$, $\kappa=0.6$, $h=0.7$ and $\beta=2.3$. Red line present $R(x,t)$ for which the integral $T(x,t)$ in Eq.~\eqref{Txt} computed exactly. Black dots present $R(x,t)$ for which we use asymptotics of integral $T(x,t)$ given by Eq.~\eqref{Txtasymp}.   
    }
    \label{FigRtime}
\end{figure}

The final formula for the asymptotics of the correlation function $G_-(x,t)$ is
\begin{multline}\label{Gtimeas}
    G_-(x,t)\approx R_\infty T(x,t)t^{-\delta_1^2-\delta_2^2}e^{iht}\\
    \times\exp \left( i\int_{-\pi}^\pi (x-t\sin q)\nu(q) dq\right),
\end{multline}
where $R_\infty$ is a constant different on each ray $v=x/t$ that additionally depends on the parameters
$\kappa$, $h$ and inverse temperature $\beta$. 
To check this asymptotics we plot in Fig.~\ref{FigRtime}  the ratio $R(x,t)$ of $G_-(x,t)$ calculated numerically from
\eqref{gmdet} to the asymptotics from the right-hand side of Eq.~\eqref{Gtimeas} without $R_\infty$.
We observe that it approaches a constant value. The possible deviations are of order $O(1/\sqrt{t})$, which is consistent 
with our approximations made for the $\nu(k)$. It would be interesting to see if these corrections can be interpreted 
in terms of the non-linear Luttinger liquid paradigm \cite{Imambekov2009,RevModPhys.84.1253}.

\section{Summary and Outlook}\label{sect:summary}

In this paper we found asymptotics of dynamical correlation functions of anyonic gas with the parameter of anyonic statistics $0\le \kappa < 1$ using recently introduced \cite{GIZ} effective form factor approach. The main difficulty of this method is to find the phase shift function $\nu(q)$ for effective fermions solving an integral equation. For large $x$ and $t$ we found approximate solutions for this integral equation which depend on the ratio $v=x/t$. For the space-like region, $v>1$, the solution $\nu(q)$ can be approximated by the smooth function $\nu_+(q)$. 
In this case the asymptotics of the correlation function is given by asymptotic analysis of integrals producing the leading contribution either from a pole or from a saddle point.
In the case of saddle-point contribution there is an additional power factor correcting the exponential decay of the correlation function.

For the time-like region, $|v|<1$, we  approximate the solution $\nu(q)$ for a large finite $t$ by a function having discontinuities at critical points
and corresponding to the solution of integral equation at $t=\infty$.
Unfortunately this approximate solution can not be used directly to find the asymptotics of correlation function 
by the methods of \cite{GIZ}, since the latter requires a smooth $\nu (q)$.
For large finite $t$ we consider a class of regularized  $\nu(q)$ having the same limit at $t=\infty$ as the genuine solution.
It is remarkable, that the regularized $\nu(q)$ lead 
to the same asymptotics up to a prefactor independent of $t$.
This universal time dependence of asymptotics has  an additional power-like factor to the exponential decay of the correlation function.  
The exponent of this power-like factor is related directly to the  jumps of  $\nu(q)$ at critical points.
We hope that the use of a better approximation to $\nu(q)$ as a solution of the integral equation for a large finite $t$ will fix the exact form of the constant prefactor. 
Further analysis of the correlation functions in the time-like region by the method of effective form factors will be presented in future publications.

The limiting case $\kappa=1$ of the model corresponds to the quantum XX spin chain model studied intensively in literature. Therefore it is interesting to look at the limits of our results as $\kappa \to 1$ 
and compare with the known formulas. For the paramagnetic phase, $h>1$, in time-like region the results for the asymptotics were obtained in \cite{XJie} 
up to an overall constant depending on $\beta$ and $h$. Our results have the same structure 
as a function of $t$.
The ferromagnetic phase,  $h<1$, was studied in \cite{Its1993,Ghmann2019a1,Ghmann2020l} in space-like region and \cite{Its1993} in time-like region.
Unfortunately, the direct application of our approach is not possible due to appearance of singularities of $\nu_\pm(q)$ at $q=\pm \arccos h$, where $\varepsilon(q)=0$.
We believe that these singularities can be properly resolved. But one needs to develop a more delicate limiting procedure, on which we hope to report in the nearest future.

An important ingredient in the derivation of asymptotics in \cite{Its1993,XJie,korepin_bogoliubov_izergin_1993}  is the use of fact that the correlation function satisfies  
differential-difference equations of Ablowitz--Ladik integrable system. It would be interesting to generalize this approach to the correlation functions with arbitrary anyonic parameter $\kappa$ and determine precise $v$ dependence of $R_\infty$ in Eq.~\eqref{Gtimeas}.

Another important application of our approach is to use it to describe the scaling behavior of the correlation functions of the anyonic gas.
One has to be able to reproduce results for the asymptotics obtained in \cite{Patu2009a,Patu2010,Santachiara2008}. Recently, using effective form factors, the finite temperature tau function for the continuum case was investigated in \cite{chernowitz2021dynamics}.

\begin{acknowledgments}
We are grateful to Pavlo Gavrylenko and  Nikita Slavnov for useful discussions.
We thank Oleg Lychkovskiy for careful reading of the manuscript and numerous useful remarks and suggestions.
The authors acknowledge support by the National Research Foundation of Ukraine grant 2020.02/0296.
Y.Z. and N.I. were partially supported by NAS of Ukraine (project No. 0117U00023).
O. G. also acknowledges support from the Polish National Agency for Academic
Exchange (NAWA) through the Grant No. PPN/ULM/2020/1/00247.

\end{acknowledgments}

\appendix*

\section{Regularization of the prefactor and power-like behaviour}
\label{appendix}

In this Appendix we describe a regularization of the divergent integral 
\begin{equation}\label{A}
    \mathcal{A}=\log Z=\frac{1}{2}\int\limits_{-\pi}^{\pi}dq\int\limits_{-\pi}^{\pi}dp\, \nu'(q)\nu'(k)\log\left|\sin\frac{q-k}{2}\right|
\end{equation}
 for the case of discontinuous $\nu(k)$.
 We use the regularization described in Eqs.~\eqref{nuR} -- \eqref{eqf1} and find asymptotics of this integral for large times.

It is natural to divide the derivative of $\nu$ into two parts
\begin{equation}
    \nu'(k)=\nu'_0(k)+\nu'_1(k),
\end{equation}
where
\begin{equation}
  \nu'_0(k)=A'(k)+B'(k)s(k), \quad \nu'_1(k)=B(k)s'(k). 
\end{equation}
In the  large $t$ limit $\nu'_0(k)$ is a bounded function meanwhile $\nu'_1(k)$ becomes proportional to $\delta$-function.
The double integral $\mathcal{A}$ can be presented as a sum of four parts
\begin{equation}
    \mathcal{A}=\mathcal{A}_{00}+\mathcal{A}_{01}+\mathcal{A}_{10}+\mathcal{A}_{11}, 
\end{equation}
where
\begin{equation}
    \mathcal{A}_{ij}=\frac{1}{2}\int\limits_{-\pi}^{\pi}dq\int\limits_{-\pi}^{\pi}dp\, 
    \nu'_i(q)\nu'_j(k)\log\left|\sin\frac{q-k}{2}\right|.
\end{equation}
Note, only $\mathcal{A}_{11}$ part is responsible for the divergence of $\mathcal A$ at large $t$.
The parts  $\mathcal{A}_{00}$,  $\mathcal{A}_{01}$ and  $\mathcal{A}_{10}$ have non-singular limiting values at $t\to\infty$ which
do not depend on regularization of $\nu(k)$. We have
\begin{equation}
    \mathcal{A}_{00}\approx\frac{1}{2}\int\limits_{-\pi}^{\pi}dq\int\limits_{-\pi}^{\pi}dp\, [\nu'](q)[\nu'](k)\log\left|\sin\frac{q-k}{2}\right| 
\end{equation}
with
\begin{equation}
[\nu'](k)=A'(k)+B'(k)\operatorname{sign}\Phi'(k).
\end{equation}
Due to $k\leftrightarrow q$ symmetry we have $\mathcal{A}_{01}=\mathcal{A}_{10}$. 
In the limit $t\to \infty$ the function $\nu'_1(k)$ becomes a sum of two delta functions and therefore
\begin{multline}
    \mathcal{A}_{01}=\mathcal{A}_{10}\approx B_1 r_1\int\limits_{-\pi}^{\pi}dq\,[\nu'](q)\log\left|\sin\frac{q-q_1}{2}\right|
    \\
    +B_2 r_2\int\limits_{-\pi}^{\pi}dq\,[\nu'](q)\log\left|\sin\frac{q-q_2}{2}\right|,
\end{multline}
where 
\begin{equation}
    B_i=B(q_i), \qquad r_i=\operatorname{sign}(\Phi''(q_i))
\end{equation}

To evaluate $\mathcal{A}_{11}$ we divide the integration region $[-\pi,\pi]$ into two pieces ${\Lambda_1=[-\pi, p]}$ and $\Lambda_2=(p, \pi]$, where point $p$ lies between critical points $q_1<p<q_2$.
This way, the double integral $\mathcal{A}_{11}$ is divided into four parts
\begin{equation}
    \mathcal{A}_{11}=a_{11}+a_{12}+a_{21}+a_{22},
\end{equation}
where
\begin{multline}
    a_{ij}=\frac{1}{2}\int\limits_{\Lambda_i}dq\int\limits_{\Lambda_j}dk\, B(q)B(k)\\
    \times s'(q)s'(k)\log\left|\sin\frac{q-k}{2}\right|.
\end{multline}
The integrals $a_{21}$ and $a_{12}$ have finite limits at $t\to\infty$
\begin{equation}
   a_{12}=a_{21}\approx 2B_1B_2r_1r_2\log\sin\frac{q_2-q_1}{2} .
\end{equation}
Remaining parts of $\mathcal{A}_{11}$ contain singularities. Let us show how they emerge on example of $a_{11}$. It is natural to present $a_{11}$ as a sum of two integrals
(regular and singular)
\begin{equation}
    a_{11}=a_{11}^{(r)}+a_{11}^{(s)},
\end{equation}
where
\begin{multline}
   a_{11}^{(r)}=\frac{1}{2}\int\limits_{\Lambda_1}dq\int\limits_{\Lambda_1}dk\, B(q)B(k)\\
   \times s'(q)s'(k)\log\left|\frac{\sin\frac{q-k}{2}}{\Phi'(q)-\Phi'(k)}\right|,
\end{multline}
\begin{multline}
   a_{11}^{(s)}=\frac{1}{2}\int\limits_{\Lambda_1}dq\int\limits_{\Lambda_1}dk\, B(q)B(k)\\
   \times s'(q)s'(k)\log\left|\Phi'(q)-\Phi'(k)\right|.
\end{multline}
The first integral can be found using L'H\^opital's rule
\begin{multline}
   a_{11}^{(r)}=\frac{2r_1\cdot 2r_1}{2}\int\limits_{\Lambda_1}dq\int\limits_{\Lambda_1}dk\, B(q)B(k)\delta(q-q_1)\delta(k-q_1)\\
   \times \log\left|\frac{\sin\frac{q-k}{2}}{\Phi'(q)-\Phi'(k)}\right|=   -2B_1^2\log|2\Phi''(q_1)|,
\end{multline}
where we used $r_1^2=1$. The second integral can be presented as 
\begin{equation}
    a_{11}^{(s)}=u_1+v_1\log\sqrt{t},
\end{equation}
where
\begin{multline}
    u_1=\frac{1}{2}\int\limits_{\Lambda_1}dq\int\limits_{\Lambda_1}dk\, B(q)B(k)\\
    \times s'(q)s'(k)\log\left|\sqrt{t}\Phi'(q)-\sqrt{t}\Phi'(k)\right|,
\end{multline}
\begin{equation}
    v_1=-\frac{1}{2}\int\limits_{\Lambda_1}dq\int\limits_{\Lambda_1}dk\, B(q)B(k)s'(q)s'(k).
\end{equation}
Performing rescaling of the integration variables one can persuade oneself that in under the last integrals $B(q)$ can be replaced to $B_1$, which leads to 
\begin{equation}
    v_1= - \frac{B_1^2}{2}\int\limits_{\Lambda_1}dq\int\limits_{\Lambda_1}dk\, s'(q)s'(k)  = - 2 B_1^2.
\end{equation}
Here we have used \eqref{eqf1}, and all the traces of the regularization has disappeared. With $u_1$ this will not be the same. 
Indeed, using $s(k)=f(\sqrt{t}\Phi'(k))$ and changing the variables of integration $q$ and $k$ by $\lambda=\sqrt{t}\Phi'(q)$ and $\mu=\sqrt{t}\Phi'(k)$ we get
\begin{equation}
    u_1=\frac{1}{2}\int\limits_{\tilde{\Lambda}_1}d\lambda\int\limits_{\tilde{\Lambda}_1}d\mu\, b(\lambda)b(\mu)f'(\lambda)f'(\mu)\log|\lambda-\mu|,
\end{equation}
where the function $b(\lambda)$ is defined as 
\begin{equation}
    b(\sqrt{t}\Phi'(q))=B(q)
\end{equation}
and region $\tilde{\Lambda}_1$ is the segment $[\sqrt{t}\Phi'(-\pi), \sqrt{t}\Phi'(p)]$ which becomes the real line when $t$ goes to infinity. 
Also $b(\lambda)$ goes to $B_1$ at $t\to \infty$. Therefore we get
\begin{equation}
    u_1\approx\frac{B_1^2}{2}\cdot \int\limits_{-\infty}^{\infty}d\lambda\int\limits_{-\infty}^{\infty}d\mu\, f'(\lambda)f'(\mu)\log|\lambda-\mu|.
\end{equation}
Finally, using  $B_1=-\delta_1/2$, $B_2=\delta_2/2$, $r_1=-1$, $r_2=1$ we obtain the following large $t$ asymptotics of $ \mathcal{A}$
\begin{equation}
   \mathcal{A}\approx d_0+d_1\log\sqrt{t},
\end{equation}
where the constant $d_1$ is universal, i.e. it is independent of regularizing function $f$
\begin{equation}
    d_1=-2(B_1^2+B_2^2)=-\frac{1}{2}(\delta_1^2+\delta_2^2)
\end{equation}
and $d_0$ depends on a regularizing function $f$ only in summands $u_1$ and $u_2$
\begin{equation}
    d_0=\mathcal{A}_{00}+2\mathcal{A}_{01}+2a_{12}+a_{11}^{(r)}+a_{22}^{(r)}+u_1+u_2.
\end{equation}

\bibliography{anyons}

\end{document}